\documentclass[aps,pre,showpacs,amsmath,amssymb,lengthcheck]{revtex4}
\usepackage{epsfig}
\usepackage{array}
\usepackage{subfigure}

\begin{document}
\newcommand{\average}[1]{\langle k^{#1} \rangle}
\newcommand{\avk}{\langle k \rangle}

\title{Biased Percolation on Scale-free Networks}

\author{Hans Hooyberghs$^{1}$, Bert Van
Schaeybroeck$^{1}$, Andr\'e A. Moreira$^{2}$, \\
Jos\'e S. Andrade, Jr.$^{2,3}$, Hans J. Herrmann$^{2,3}$ and
Joseph O. Indekeu$^{1}$}

\affiliation{$^{1}$Instituut voor Theoretische Fysica, Katholieke
Universiteit Leuven, Celestijnenlaan 200 D, B-3001 Leuven,
Belgium\\
$^{2}$Departamento de F\'isica, Universidade Federal do Cear\'a,
60451-970 Fortaleza, Cear\'a, Brazil\\
$^{3}$Computational Physics, IfB, ETH-H\"onggerberg,
Schafmattstrasse 6, CH-8093 Z\"urich, Switzerland}
\date{\small\it \today}

\begin{abstract}
Biased (degree-dependent) percolation was recently shown to
provide new strategies for turning robust networks fragile and
vice versa. Here we present more detailed results for biased edge
percolation on scale-free networks. We assume a network in which
the probability for an edge between nodes $i$ and $j$ to be
retained is proportional to $(k_ik_j)^{-\alpha}$ with $k_i$ and
$k_j$ the degrees of the nodes.  We discuss two methods of network
reconstruction, sequential and simultaneous, and investigate their
properties by analytical and numerical means. The system is
examined away from the percolation transition, where the size of
the giant cluster is obtained, and close to the transition, where
nonuniversal critical exponents are extracted using the generating
functions method. The theory is found to agree quite well with
simulations. By introducing an extension of the Fortuin-Kasteleyn
construction, we find that biased percolation is well described by
the $q\rightarrow 1$ limit of the $q$-state Potts model with
inhomogeneous couplings.
\end{abstract}


\maketitle

\section{Introduction}
In recent years, much attention has been devoted to the study of
real-life networks. Such networks may be modelled by points or
nodes connected by edges. One feature is the scale-free topology,
described by a probability distribution $P(k)$ for the number of
edges $k$ of a node, which falls off as a power law $k^{-\gamma}$
for large values of $k$. Most of the investigated cases turn out
to have a topological exponent, or ``degree exponent", $\gamma$,
in the range $2 < \gamma < 3.5$. For $\gamma
> 2$ the mean degree $\langle k \rangle$ is finite, and for $\gamma > 3$ also the
variance $\langle k^2\rangle$ is finite. Like fully random
Poisson-distributed networks (with a typical scale), also
scale-free networks are of ``small-world" type. By now, many
properties have been revealed and investigated thoroughly: these
include degree-degree correlations, clustering and directedness of
the edges in the
network~\cite{albert,boccalettia,dorogovtsev,dorogovtsev2} .

Another well-known property of scale-free networks is their
resilience against random failure, a robustness caused by the
presence of hubs (nodes with very high degree). On the other hand,
these hubs may cause the network to be very vulnerable when a
targeted attack is performed. In the limit of infinitely large
networks, the network is said to be \textit{robust} when even
after removing an arbitrary fraction of the edges, there is still
a nonzero probability that two randomly chosen nodes are part of a
connected cluster. On the other hand, when removing edges from a
\textit{fragile} network, a point will be reached when the giant
cluster, the one with a size comparable to the network size, is
destroyed; this very point is called the percolation threshold.
The percolation transition is a genuine phase transition and is
normally of second order so that critical exponents can be
properly defined~\cite{goltsev,schwartz,cohen3}.

Since the first studies of percolation on scale-free
networks~\cite{cohen}, a lot of work has been done on node
percolation~\cite{callaway,cohen,cohen2,newman3,gallos,cohen3},
bond percolation~\cite{newman,callaway,allard,dallasta},
percolation on multitype
networks~\cite{allard,newman,newman2,dallasta}, clustered
networks~\cite{serrano2,serrano3}, correlated
networks~\cite{newman3,goltsev,serrano3,serrano4}, directed
networks~\cite{schwartz,serrano4,newman2}, degree-dependent edge
percolation~\cite{newman,dallasta,wu} and degree-dependent node
percolation~\cite{gallos}.

The percolation transition has many connections to real systems.
For example, it can be related to disease propagation
models~\cite{newman,kenah,hastings},~\footnote{In
Ref.~\cite{newman}, the author argued that a large class of the
standard epidemiological or disease propagation models can be
exactly mapped onto percolation. However, this claim was disproven
in Refs.~\onlinecite{kenah} and~\onlinecite{hastings}. The authors
of Ref.~\onlinecite{kenah}, however, showed that a connection to a
more involved percolation problem is valid.}. In this analogy, the
infection of an individual is represented by the activation of a
node of the (social) network. When a giant cluster of active nodes
emerges, an epidemic is established. Disease propagation on such
networks can be efficiently suppressed by selective vaccination,
depending for example on the connectedness of each node.

An alternative interpretation of a network with a certain fraction
of deactivated edges is in terms of a transport network in which
the edges transmit data or deliverables between nodes with a
certain transmission probability. This probability depends in
general on the degrees of the connected nodes. For example,
communication with the highly connected hubs on the internet is in
general more efficient. It is, however, also possible that nodes
with more edges are less robust. Indeed, in more social terms,
friendships involving people which have many acquaintances are
more likely to end than friendships between people with few
connections. Or, as another example, traffic on a network induces
high loads on highly connected nodes which in turn makes them more
vulnerable to failure. Clearly, the resilience of an edge in a
real network may depend strongly on the degrees of the nodes it
connects.

We study the properties of a network after biased or
degree-dependent edge removal. More specifically, we consider
networks in which the edge between nodes $i$ and $j$ is {\em
retained} with a probability proportional to its weight
\begin{align}\label{weights}
w_{ij}=(k_ik_j)^{-\alpha},
\end{align}
where $k_i$ and $k_j$ denote the degrees of nodes $i$ and $j$
respectively, and $\alpha$ is the ``bias exponent''. By tuning
$\alpha$, we can explore three qualitatively different regimes:
random failure ($\alpha=0$), the attack of edges connected to hubs
($\alpha>0$), and the depreciation of edges between the least
connected nodes ($\alpha<0$). Henceforth, we call the regime
$\alpha>0$ ``centrally biased'' (CB). The converse regime,
$\alpha<0$, is termed ``peripherally biased'' (PB).

A degree dependence similar to that in Eq.~\eqref{weights} has
already been considered in Refs.~\onlinecite{giuraniuc1}
and~\onlinecite{giuraniuc2} where Ising spin couplings $J_{ij}$ on
scale-free networks were taken to be proportional to $w_{ij}$. The
motivation for introducing degree-dependent couplings was the
observation that for $\gamma \le 3$ the system is always
``ordered" (critical temperature $T_c = \infty$) due to the
dominance of the hubs. However, degree-dependent couplings make it
possible to compensate high degree with weak interaction (assuming
$\alpha
>0$) so that the effect of the hubs can be neutralized.
In doing so, it was discovered that a network with ``interaction
exponent" $\alpha$ and degree exponent $\gamma$ has the same
critical behavior as a network with interaction exponent zero
(uniform couplings $J$) and degree exponent
\begin{align}\label{mappy}
\overline{\gamma}=\frac{\gamma-\alpha}{1-\alpha}.
\end{align}
In this way it was possible to ``trade interactions for topology"
and study the rich mean-field critical behaviour, with
nonuniversal critical exponents depending on $\gamma$
\cite{dorogovtsev2}, simply by varying $\alpha$ in a given network
with fixed $\gamma$. The same exponent mapping will be recovered
in this work in the following sense: at percolation the properties
of a network with bias exponent $\alpha$ and degree exponent
$\gamma$ are the same as those of a network with bias exponent
zero and degree exponent $\overline{\gamma}$, {\em or} degree
exponent $\gamma$, depending on conditions that will be specified.

The significance and potential usefulness of biased depreciation
of a network is now becoming more clear. Indeed, it has been shown
that networks with $\gamma > 3$ are {\em fragile} under random
failure, while networks with $\gamma < 3$ are {\em robust} under
random removal of edges or nodes \cite{cohen}. If it should turn
out, and under certain conditions this is what we find, that the
depreciated network behaves as one in which $\gamma$ is replaced
by $\overline{\gamma}$, it becomes possible to control the
robustness or fragility of a network systematically by tuning the
bias exponent $\alpha$. In other words, a network that is robust
under random failure may turn out to be fragile under biased
failure, and the other way round. Note that applying bias does not
presuppose global knowledge about the network (location of the
hubs, ...) but only requires local information on nodes and their
degree.

The exponent equality can be intuitively understood from the
following heuristic argument, which is safe to use provided
$\alpha > 0$ and $k$ is sufficiently large. Using
Eq.~\eqref{weights}, one can anticipate that after depreciation of
the network, a node with degree $k$ will, on average, have a new
degree $\overline{k}$ proportional to $k^{1-\alpha}$. Since all
nodes remain in place during the depreciation process, the
original degree distribution $P(k)$ changes into a new
distribution $\overline{P}(\overline{k})$ after depreciation, the
relation between them being:
\begin{align}\label{mappy}
P(k)dk=\overline{P}(\overline{k})d\overline{k}.
\end{align}
Using $\overline{k}\propto k^{1-\alpha}$, one directly infers that
indeed $\overline{P}(\overline{k})\propto
\overline{k}^{-\overline{\gamma}}$ and the network after
depreciation thus acquires degree exponent $\overline{\gamma}$ and
the corresponding percolation properties. A more rigorous proof of
this plausible expectation is given in the Appendix.

The paper is organized as follows: In
Secs.~\ref{sequentialapproach} and~\ref{simultaneousapproach} we
introduce random scale-free networks and present two distinct
approaches by means of which a network can be reconstructed in a
degree-dependent manner. Based on these schemes, we focus in
Sec.~\ref{characteristics} on the degree distribution and the
degree-degree correlations of the network after (partial)
reconstruction. The percolation threshold is then extracted from
these degree characteristics in Sec.~\ref{percolationtrigger}. The
theory of generating functions for degree-dependent percolation on
random networks will be extensively presented in
Sec.~\ref{generousfunctions}. In Sec.~\ref{kastfort} the
equivalence of our model with the Potts model is elaborated and
using this equivalence and finite-size scaling theory, we arrive
at the critical exponents for the percolation transition in
Sec.~\ref{critexponents}. Finally, our results are extensively
compared to simulational results in Sec.~\ref{comparison}. Our
conclusions are presented in Sect.~\ref{conclusions}. A summary of
part of the results presented here has been reported in
Ref.~\onlinecite{moreira}.

\section{Degree-Dependent Percolation On Random
Graphs}\label{sec_intro}

This section concerns (maximally) random scale-free networks.
These are networks generated with the so-called configuration
model, which assumes that the degrees of the nodes in the network
are distributed according to a probability $P(k)$ which is taken
to be the power law:
\begin{align}\label{nodedistrib}
P(k) = C k^{-\gamma},
\end{align}
for values of $k$ between the minimal and maximal degrees $m$ and
$K$, respectively. $C$ is the normalization constant. In order to
ensure a finite mean degree we take $\gamma > 2$. The graph is
then completed by connecting the stubs emanating from all nodes.
The probability $P_n(k)$ that a randomly chosen edge leads to a
node of degree $k$ must therefore be:
\begin{align}\label{nearestn}
P_n(k) = \frac{kP(k)}{\avk},
\end{align}
where $\langle \cdot \rangle$ denotes the average over the nodes,
obtained using probability distribution $P(k)$. The probability
distribution $P_n$ is also called the nearest-neighbor degree
distribution. \textit{Random networks} are constructed by
connecting the earlier mentioned stubs randomly. We do not allow
self-connections nor multiple connections between nodes and use
the method proposed in Ref.~\onlinecite{catanzaro} to avoid
degree-degree correlations in the network. To quantify
degree-degree correlations, let us introduce the probability
$P(k,q)$ that nodes of degree $k$ and $q$ are connected. If no
correlations are present, $P(k,q)$ reduces to:
\begin{align}
P(k,q)=P_n(k)P_n(q)=\frac{kqP(k)P(q)}{\avk^2}.
\end{align}
Below and close to the critical point, large and random networks
can locally be treated as trees and loops are sparse so that their
effect can, to a good approximation, be ignored. The local
tree-like structure will be used in Sec.~\ref{generousfunctions}
when the generating functions method is introduced.

We continue with presenting two distinct depreciation methods to
study degree-dependent percolation. The statistical edge
properties are now being considered, which will allow us in
Sec.~\ref{characteristics} to obtain the statistics of nodal
properties.

\subsection{Sequential Approach}\label{sequentialapproach}
Our first method, which we call the \textit{sequential approach},
starts from a random network with all $N$ node degrees distributed
according to the degree distribution $P(k)$. Initially all edges
are removed and we aim at reintroducing a fraction $f$ of the
total number of edges $N_e=\langle k\rangle N/2$. This is achieved
by activating one edge in each time step $t$. Consequently, the
probability that the edge between nodes $i$ and $j$ is activated
is $w_{ij}/Z_t$ where $Z_t$ is the sum of weights $w_{ij}$ of all
non-activated edges after $t-1$ steps. Thus, the probability
$\rho_{ij}(f)$ that an edge between nodes $i$ and $j$ is again
present after the reinclusion of a fraction $f$ of the edges,
is~\cite{cohen4}:
\begin{align}\label{john}
\rho_{ij}(f) = 1 - \prod_{t=1}^{fN_e} \left(1 - \frac{w_{ij}}{Z_t}
\right).
\end{align}
For sufficiently large networks $w_{ij}/ Z_t$ is typically small
compared to one and Eq.~\eqref{john} is well approximated by:
\begin{align}\label{timothy}
\rho_{ij}(f) \approx 1 - e^{-D_fw_{ij}},
\end{align}
with the positive parameter $D_f=\sum_{t=1}^{fN_e} Z_t^{-1}$. It
can be argued that for a sufficiently narrow distribution of the
weights~\footnote{This is valid as long as $t$ is small enough so
that $(t-1)(\langle w^2\rangle_e-\langle w\rangle_e^2)/\langle
w\rangle_e^2\ll N_e$.}:
\begin{align}\label{justlinear}
Z_t=\langle w\rangle_e(N_e -t+1),
\end{align}
where $\langle\cdot \rangle_e$ denotes the average over all edges.
The following property is readily derived,
\begin{align}\label{wproperty}
\sqrt{\langle w\rangle_e} = \frac{\langle
k^{1-\alpha}\rangle}{\langle k \rangle} = \frac{\gamma -2}{\gamma
-2 + \alpha} m^{-\alpha}.
\end{align}

Using Eq.~\eqref{justlinear}, $D_f$ can be determined, such that
for large $N_e$,
\begin{align}\label{dfexplicit}
D_f=-\ln\left[1-f\right]/\langle w\rangle_e
\end{align}
and
thus~\cite{abramowitz}:
\begin{align}\label{snoopy}
\rho_{ij}(f) =1-\left[1-f\right]^{w_{ij}/\langle w\rangle_e}
\end{align}
It is instructive to consider a few asymptotic regimes of
Eq.~\eqref{snoopy}. First, in the case $\alpha=0$, one recovers
the expression for degree-independent percolation $\rho_{ij}=f$ as
expected. Secondly, for arbitrary $\alpha$, we can distinguish the
dilute limit and the dense limit in terms of $f$, and find:
\begin{subequations}
\begin{align}
\rho_{ij}&\sim fw_{ij}/\langle w\rangle_e\text{ when }f \rightarrow 0,\label{tim}\\
\rho_{ij}& \sim 1\text{ when }f \rightarrow 1 .
\end{align}
\end{subequations}
Thirdly, when $w_{ij}/\langle w\rangle_e\ll
[-\ln\left(1-f\right)]^{-1}$:
\begin{align}\label{beammeup}
\rho_{ij}\sim -\ln\left(1-f\right)w_{ij}/\langle w\rangle_e.
\end{align}

We proceed by defining the marginal distribution $\rho_k$ as the
mean probability that an edge connected to a node with degree $k$
is present in the network after reconstruction. Thus
\begin{align}\label{marginal}
\rho_k= \sum_{q=m}^K P_n(q)\rho_{kq},
\end{align}
where $\rho_{kq}=1-e^{-D_fw_{kq}}$. A good analytic approximation
to $\rho_k$ can be obtained by substituting $\rho_{kq}$ into
Eq.~\eqref{marginal}, considering an integral instead of a sum,
taking the macroscopic limit $K \rightarrow \infty$, and expanding
the exponential,
\begin{align}
\rho_k = 1 - \sum^{\infty}_{n=0}\frac{(-)^n (D_f \,m^{-\alpha}
k^{-\alpha})^n}{(1+ n\alpha/(\gamma-2))\,n!}.
\end{align}
Alternatively, this result follows straightforwardly from the fact
that the marginal distribution involves the incomplete Gamma
function \cite{abramowitz}. The usefulness of this explicit form
can best be appreciated by first considering the range $0<\alpha
\ll \gamma -2$, for which we obtain the simple analytic result
\begin{eqnarray}\label{ghinzu}
\rho_k &\sim & 1 - \exp\left\{- \frac{\gamma -2}{\gamma -2 +
\alpha} D_f m^{-\alpha} k^{-\alpha}\right\}\nonumber \\
& =& 1 - \exp\left\{- D_f \sqrt{\langle
w\rangle_e}\,k^{-\alpha}\right\}.
\end{eqnarray}
Although this result is strictly only valid for the specified
range of $\alpha$ specified above, numerical inspection shows that
it is a rather good approximation to the integral representation
of the sum Eq.~\eqref{marginal} for a wider range of $\alpha$,
{\em including negative values}. In fact, the result is useful in
the entire interval of our interest $\alpha \in [2-\gamma, 1]$.
Using the previously obtained approximation to $D_f$,
Eq.~\eqref{dfexplicit}, it can be further simplified to
\begin{align}\label{statement}
\rho_k \approx 1 - [1-f]^{k^{-\alpha}/\sqrt{\langle w\rangle_e}}.
\end{align}

Based on the asymptotic regimes of $\rho_{ij}$ it is also possible
to extract the behavior of $\rho_k$. When $f\rightarrow 0$, we may
use Eq.~\eqref{tim}:
\begin{align}\label{bob}
\rho_{k}&\sim k^{-\alpha}f/\sqrt{\langle w\rangle_e}
\end{align}
On the other hand, when $\alpha>0$ and $k\gg k_{\times}$, where
the cross-over value for $k$ is given by $k_{\times} \equiv
D_f^{1/\alpha}/m$, we get:
\begin{align}\label{scalubl}
\rho_{k}\sim k^{-\alpha}D_f \sqrt{\langle w\rangle_e}\approx
-k^{-\alpha}\ln\left(1-f\right)/ \sqrt{\langle w\rangle_e}.
\end{align}

\subsection{Simultaneous Approach}\label{simultaneousapproach}
As an alternative to the sequential approach we introduce now the
simultaneous approach. Again we start from a fully depreciated
uncorrelated network with degree distribution $P(k)$. We then
visit each edge (between nodes $i$ and $j$) once and activate this
edge with probability
\begin{align}\label{simultaneousexpression}
\rho_{ij}=f w_{ij}/\langle w\rangle_e.
\end{align}
In contrast to the sequential approach, $\rho_{ij}$ is now
history-independent. Note also that $\langle \rho_{ij}\rangle_e=f$
as it must be. For the marginal distribution $\rho_{k}$ one finds:
\begin{align}\label{marginalpaparel}
\rho_{k}= k^{-\alpha}f/\sqrt{\langle w\rangle_e},
\end{align}
which satisfies $f\rho_{kq}=\rho_{k}\rho_{q}$ and is the same as
in the $f\rightarrow 0$ limit of the sequential approach
(Eq.~\eqref{bob}). However, for each value of $k$ and $q$, the
probability $\rho_{kq}$ must be less than or equal to one. This
means that Eq.~\eqref{simultaneousexpression} is only well-defined
for values of $f$ for which
\begin{align}\label{fu}
f< f_u\equiv\left(\frac{\langle k^{1-\alpha}\rangle}{\langle
k\rangle}\right)^2\times (\text{min}(m^{\alpha},K^{\alpha}))^2.
\end{align}
It can be calculated that, in the macroscopic limit, the rhs of
Eq.~\eqref{fu} vanishes when $\alpha<0$ and therefore the
simultaneous approach is only meaningful for positive $\alpha$ and
provided
\begin{align}\label{fff}
f<\left(\frac{\gamma-2}{\gamma-2+\alpha}\right)^2.
\end{align}
To reach fractions above this limit in the simulations, we {\em
iterate} the simultaneous approach. The first iteration involves
the usual simultaneous approach with $f=f_u$; the second iteration
is initialized by considering a new network consisting of all
edges that have not been reintroduced during the first sweep. For
that network one calculates the probabilities $w_{ij}/Z_2$ and a
new value of $f_u$, which is the minimum of the set $\{\langle
w\rangle_{e}/w_{ij}\}$ where also the average is only over edges
of the new network. One then applies the simultaneous approach
until the new $f_u$ is reached, after which a third iteration can
be initialized if necessary. Such iterations, however, introduce
correlations and history dependence. Note that the sequential
approach can be seen as an extreme case of an iterated
simultaneous approach in which only one edge is reconstructed in
each iteration.

\subsection{Degree Distribution and Correlations of the Reconstructed Network}\label{characteristics}

We now seek to obtain the degree distribution and characterize
degree-degree correlations for the network after reconstruction.
The following is valid for both the simultaneous and sequential
approaches. Henceforth we adopt the convention that an overbar
indicates quantities in the diluted, or depreciated, network.

For the node degree distribution $\overline{P}(\overline{k})$ and
the degree-degree correlations embodied in
$\overline{P}(\overline{k},\overline{q})$ of the depreciated
network we can write
\begin{subequations}
\begin{align}
\overline{P}(\overline{k})&=\sum_{k=\overline{k}}^KP(A_{k}\wedge
B_{k\rightarrow \overline{k}}),\\
\overline{P}(\overline{k},\overline{q})&=\sum_{k=\overline{k}}^K\sum_{q=\overline{q}}^KP(C_{qk}\wedge
B_{q\rightarrow \overline{q}}\wedge B_{k\rightarrow
\overline{k}}\wedge D)/f.\label{fritz}
\end{align}
\end{subequations}
Here we introduced the notation, for events A-D,
\begin{itemize}

\item $A_{k}$: a randomly chosen node of the \textit{original}
network has degree $k$.

\item $B_{k\rightarrow \overline{k}}$: the degree of a node goes
from $k$ in the \textit{original} to $\overline{k}$ in the
\textit{depreciated} network.

\item $C_{qk}$: the nodes connected by a randomly chosen edge of
the \textit{original} network have degrees $q$ and $k$.

\item $D$: the chosen edge has not been removed from the
\textit{original} network.

\end{itemize}
For the node degree distribution, one readily finds
\begin{align}\label{binomial}
\overline{P}(\overline{k})&= \sum_{k=\overline{k}}^K P(k)\left(
\begin{array}{c}
k \\
\overline{k}
\end{array} \right)
\rho_{k}^{\overline{k}}(1-\rho_{k})^{k-\overline{k}}.
\end{align}
For large values of $k$, i.e., $k \gg k_{\times}$, and $\alpha>0$,
the probability of retaining a node of degree $k$ falls off as
$\rho_k\propto k^{-\alpha}$; this is valid using the sequential
approach (see Eq.~\eqref{scalubl}), as well as the simultaneous
one (see Eq.~\eqref{marginalpaparel}). Substituting this into
Eq.~\eqref{binomial} and approximating the binomial distribution
in Eq.~\eqref{binomial} by a normal distribution, one arrives at:
\begin{align}\label{propto}
\overline{P}(\overline{k})\propto
\overline{k}^{-\overline{\gamma}} \text{ for }
\overline{k}\rightarrow \infty.
\end{align}
where $\overline{\gamma}$ is defined in Eq.~\eqref{mappy}. This
result, which is proven in the Appendix, confirms the validity of
the expectation raised in the Introduction. Note that in case
$\alpha=0$, $\overline{P}(\overline{k})\propto
\overline{k}^{-\gamma}$ as it must be. We introduce now averaging
over nodes of the reconstructed network:
\begin{align}
\ll \cdot\gg=\sum_{\overline{k}=0}^K
\overline{P}(\overline{k})\,\cdot.
\end{align}
For further purposes, we calculate now the first and second moment
of $\overline{P}(\overline{k})$ in terms of the moments of $P(k)$:
\begin{subequations}\label{newmoments}
\begin{align}
\ll \overline{k}\gg&=\langle k\rho_k\rangle=f\langle k\rangle,\\
\ll \overline{k}^2\gg&=\langle k\rho_k(k\rho_k-\rho_k+1)\rangle.
\end{align}
\end{subequations}

The degree-degree correlations are embodied in Eq.~\eqref{fritz}.
This function can be further worked out to yield:
\begin{align}
\overline{P}(\overline{k},\overline{q})=&\sum_{k=\overline{k}}^K\sum_{q=\overline{q}}^KP_n(q)P_n(k)\rho_{kq}\nonumber\\
&\times\left(
\begin{array}{c}
q-1\\
\overline{q}-1
\end{array} \right)
\rho_{q}^{\overline{q}-1}(1-\rho_{q})^{q-\overline{q}}\nonumber\\
&\times \left(
\begin{array}{c}
k-1\\
\overline{k}-1
\end{array} \right)
\rho_{k}^{\overline{k}-1}(1-\rho_{k})^{k-\overline{k}}/f.
\end{align}
This can be reduced to:
\begin{align}\label{correlated}
\overline{P}(\overline{k},\overline{q})=&\overline{k}\overline{q}\sum_{k=\overline{k}}^K\sum_{q=
\overline{q}}^K\frac{P(q)P(k)}{(f\langle k\rangle)^2}
P(B_{q\rightarrow \overline{q}})P(B_{k\rightarrow
\overline{k}})\frac{f\rho_{kq}}{\rho_{k}\rho_{q}}.
\end{align}
Eq.~\eqref{correlated} expresses the degree-degree correlations of
a network after degree-dependent depreciation of a fully
uncorrelated network. The question that can now be raised is when
the depreciated network is also free of correlations, or, when is
$
\overline{P}(\overline{k},\overline{q})=\overline{P}_n(\overline{k})\overline{P}_n(\overline{q})$?
It is readily checked that this is true provided
\begin{align}\label{correlationfreecriterion}
f\rho_{kq}=\rho_{k}\rho_{q}.
\end{align}
As Eq.~\eqref{correlationfreecriterion} is valid for the
simultaneous approach (see Eqs.~\eqref{marginalpaparel}
and~\eqref{simultaneousexpression}), no correlations appear in the
reconstructed network (after a single iteration). For the
sequential approach, on the other hand,
Eq.~\eqref{correlationfreecriterion} is generally not satisfied
and the reconstructed network will be correlated.

The following limiting case of the \textit{sequential} approach is
interesting: take $\alpha$ positive and consider an edge between
two nodes of large degrees $k$ and $q$ such that $k\gg k_{\times}
$ and $q\gg k_{\times}$. We may then substitute
Eqs.~\eqref{beammeup} and~\eqref{scalubl} into
Eq.~\eqref{correlated}. One soon arrives at the result:
\begin{align}\label{disassort}
\frac{\overline{P}(\overline{k},\overline{q})}{\overline{P}_n(\overline{k})\overline{P}_n(\overline{q})}=-\frac{f}{\ln\left(1-f\right)}.
\end{align}
Since the rhs is smaller than one, this demonstrates that the
sequential approach causes \textit{disassortative mixing} in the
depreciated network when $\alpha>0$. In other words, nodes with
large degrees tend to be connected to nodes with small degrees and
vice versa. Using simulations, we will present evidence in
Sect.~\ref{comparison} that such correlations are introduced.

Finally, note also that Eq.~\eqref{correlated} reduces to the
correct nearest-neighbor degree distribution upon summing over
$\overline{q}$:
\begin{align}\label{nearestneighbor}
\overline{P}_n(\overline{k})=\frac{\overline{k}\,\overline{P}(\overline{k})}{f\langle
k\rangle}.
\end{align}

\section{Percolation Threshold for central bias}\label{percolationtrigger}
Here we focus solely on centrally biased (CB) depreciation
($\alpha>0$) using the simultaneous approach. One may wonder what
happens if centrally biased (CB) depreciation is applied to a
robust network with $\gamma<3$ such that the edges between and
emanating from hubs are preferentially removed. Since, in that
case, $\overline{\gamma}
> \gamma$, one may speculate that a robust network may turn
fragile and that the threshold for this to occur is
$\overline{\gamma}=3$ instead of the threshold $\gamma=3$ valid
for degree-independent percolation. We will address this question
further and conclude that it is indeed so.

On the other hand, if we start from a fragile network ($\gamma
>3$) and apply CB, it is logical that the net remains fragile.
Upon removing edges linked to hubs with a larger probability, we
are more likely to destroy the coherence of the network. The
question can then still be posed how much the percolation
threshold of the reconstructed network is shifted.

Our first task now is to calculate the critical fraction at which
the network becomes disconnected. According to Molloy and
Reed~\cite{molloy}, the critical fraction of a \textit{random}
network can be found by looking at the average nearest-neighbor
distribution. If, upon following a random edge, the attained node
has more than two neighbors, the network is said to be
percolating, that is, a giant cluster will be present in the
network. Note that this criterion is exact for the simultaneous
but not for the sequential approach, due to the appearance of
degree-degree correlations. In the reconstructed network, the
Molloy-Reed criterion reads:
\begin{align}\label{molloyreed}
1=\frac{\ll \overline{k}(\overline{k}-1)\gg}{\ll \overline{k}\gg},
\end{align}
or equivalently, using Eq.~\eqref{newmoments}:
\begin{align}\label{starsturntodust}
2\langle k\rho_k\rangle=\langle k\rho_k(k\rho_k-\rho_k+1)\rangle.
\end{align}
Using Eq.~\eqref{marginalpaparel} for the simultaneous approach,
we find the following expression for the critical fraction $f_c$
at percolation~\cite{moreira}:
\begin{align}\label{criterion584}
f_c=\frac{\langle k^{1-\alpha}\rangle^2}{\langle k\rangle(\langle
k^{2-2\alpha}\rangle-\langle k^{1-2\alpha}\rangle)}.
\end{align}
For unbiased depreciation of the network ($\alpha=0$) this last
expression reduces to the well-known formula of random percolation
on random networks~\cite{cohen}.

Eq.~\eqref{criterion584} allows us now to find out whether the
network is robust, or in other words, whether $f_c\rightarrow 0$.
This vanishing occurs, in the macroscopic limit, when the term
$\langle k^{2-2\alpha}\rangle$ diverges and therefore:
\begin{align}
\overline{\gamma}
\begin{cases}
>3\text{:  the network is fragile,} \\
<3\text{:  the network is robust}.
\end{cases}
\end{align}
The scaling relation of $f_c$ as a function of the network size
can be found when $\overline{\gamma} <3$; using $N\propto
K^{\gamma-1}$, one finds:
\begin{align}
f_c \propto N^{\frac{\overline{\gamma}-3}{\overline{\gamma}-1}}.
\label{fcN}
\end{align}

Whether or not a network is robust for the degree-dependent attack
is thus not solely a property of the network. Also the exponent
$\alpha$ plays a crucial role in the arguments and its effect can
be absorbed by using the exponent $\overline{\gamma}$ instead of
the exponent before dilution, $\gamma$. In
Sect.~\ref{critexponents} we will take a closer look at the regime
around the percolation threshold and we will find that the same
mapping from $\gamma$ to $\overline{\gamma}$ is valid.

\section{Generating Functions Approach}\label{generousfunctions}
We now introduce the generating functions approach for
degree-dependent percolation. By this method, certain properties
of the finite clusters in the network are easily obtained; this in
turn allows to draw conclusions about the giant cluster. The
method is exact if loops in the network can be ignored; since in
the macroscopic limit, the average loop sizes in the finite
clusters diverge~\cite{bianconi}, the method turns out to be
exact. This will be apparent in Sect.~\ref{comparison} when
comparing the analytical results with simulations.

\subsection{Introduction}\label{randomnetworks}
Generating functions are used in a wide branch of mathematical
problems concerning series \cite{wilf}. A generating function of a
series is the power series which has as coefficients the elements
of the series. Applied to the context of percolation problems,
this series is taken to be that of the discrete probability
distributions characterizing the network under
consideration~\cite{essam,newman}. We explain first the general
formalism while closely following the approach of
Newman~\cite{newman}, which we adapt for degree-dependent edge
percolation~\footnote{Note that in this context, the generating
functions approach is an exact method, except for the neglect of
loops in the network. Anyway, the amount of loops in the finite
clusters vanishes in the macroscopic limit.}.

The most fundamental generating function is the one that generates
the degree distribution of the network
\begin{align}
G_0(h) = \sum_{k=m}^K P(k) e^{-hk}.\label{G0}
\end{align}
We also define the generating function for the distribution of
residual edges of a node reached upon following a random edge:
\begin{align}
G_1(h) = \sum_{k=m}^K P_n(k)e^{-h(k-1)}. \label{G0}
\end{align}
The exponent of $e^{-h}$ is $k-1$ because the edge which is used
to reach the node is not counted.

There is a threefold advantage in working with the generating
functions instead of working with the degree distribution itself.
First, moments can be obtained easily from the generating
functions. For instance, the average degree is given by
\begin{align}
\avk = \sum_{k=m}^K kP(k) = -G_0'(0),
\end{align}
where $G_0'$ denotes the derivative with respect to $h$. Higher
moments can be obtained with higher-order derivatives.

Secondly, we can benefit from the so-called \textit{powers
property} of generating functions: if the distribution of a
property $k$ of an object is generated by a function $G(h)$, then
the generating function of the sum of $n$ independent realizations
of $k$ is $G(h)^n$. For instance, if we randomly choose $n$ nodes
in our network, the distribution of the sum of the degrees is
generated by $G_0(h)^n$.

Thirdly, the use of generating functions will allow us in
Sect.~\ref{kastfort} to highlight the equivalence with the
$q\rightarrow 1$ limit of the $q$-state Potts model where the
parameter $h$ will play the role of the magnetic field.

\subsection{Self-consistent equations}\label{selfconsistent}
We can now define the equivalents of $G_0(h)$ and $G_1(h)$ for the
network \textit{after dilution} as $F_0(h)$ and $F_1(h)$:
\begin{subequations}\label{genererendefunctiesrule}
\begin{align}
F_0(h)&=\sum_{\overline{k}=0}^K \overline{P}(\overline{k}) e^{-h\overline{k}},\\
F_1(h)&=\sum_{\overline{k}=0}^K
\overline{P}_n(\overline{k})e^{-h(\overline{k}-1)}.
\end{align}
\end{subequations}
Note that the minimal degree in the network after dilution is zero
instead of $m$. Eq.~\eqref{genererendefunctiesrule} can be worked
out further using Eqs.~\eqref{binomial}
and~\eqref{nearestneighbor}:
\begin{subequations}
\begin{align}
F_0(h) &= \sum_{k=m}^K P(k) (1-\rho_k + e^{-h}\rho_k)^k\label{F0},\\
F_1(h) &= \sum_{k=m}^K \frac{\rho_k P_n(k) }{f}(1-\rho_k +
e^{-h}\rho_k)^{k-1} \label{F1}.
\end{align}
\end{subequations}

The most interesting quantity for us is the size distribution of
the finite clusters, the generating function of which can be
readily derived using $F_0$ and $F_1$. Let $H_0$ denote the
generating function for the probability that a randomly chosen
node belongs to a connected cluster of a given (finite) size.
Furthermore, let $H_1$ be the generating function for the
probability that upon following a randomly chosen edge to one end,
a cluster of a given (finite) size is reached. If the network can
be treated as a tree, these generating functions satisfy the
following self-consistency equations~\footnote{As explained later,
these self-consistency equations are only valid in case the
simultaneous approach is used, that is when no correlations are
present in the network after depreciation.}:
\begin{subequations}\label{H}
\begin{eqnarray}
H_1(h) &=& e^{-h}F_1[H_1(h)],\label{H0}\\
H_0(h) &=& e^{-h}F_0[H_1(h)].\label{H1}
\end{eqnarray}
\end{subequations}
Here the function $F_{0,1}[H_1(h)]$ denotes the function $F_{0,1}$
in which $e^{-h}$ is replaced by $H_1(h)$. The proof of these
relations relies on the aforementioned powers property and is
expounded in Ref.~\onlinecite{newman}. The percolation threshold
can now be derived with the aid of these functions.

Several macroscopic quantities can be easily identified in the
\textit{depreciated} network~\cite{dorogovtsev2}. For example, we
define $\mathcal{P}_{\infty}$ as the probability that a node
belongs to the giant cluster, $\mathcal{L}_{\infty}$ as the edge
probability for being in the giant cluster and $\mathcal{S}$ as
the average cluster size of finite clusters:
\begin{subequations}\label{richard}
\begin{align}
&\mathcal{P}_{\infty}=1-H_0(0),\\
&\mathcal{L}_{\infty}=1-(H_1(0))^2,\\
&\mathcal{S}=-H_0'(0).\label{timothy2}
\end{align}
\end{subequations}
Moreover, the degree distribution of nodes in the giant cluster
varies as
\begin{align}
\overline{P}_{gc}(\overline{k})\propto (1-(H_1(0))^{\overline{k}}
)\overline{P}(\overline{k}),
\end{align}
from which it follows that the degree distribution of the finite
clusters varies as $\overline{P}_{fc}(\overline{k})\propto
(H_1(0))^{\overline{k}} \overline{P}(\overline{k})$. In case there
are both finite clusters and a giant cluster, the degree
distributions have the asymptotic behavior
($\overline{k}\rightarrow \infty$):
\begin{subequations}
\begin{align}
&\overline{P}_{gc}(\overline{k})\sim \overline k\,^{-\overline{\gamma}},\\
&\overline{P}_{fc}(\overline{k})\sim e^{-\overline{k}/\lambda},
\end{align}
\end{subequations}
with $\lambda=-\ln(H_1(0))$. In other words, in the presence of a
giant cluster, only the degree distribution of the giant cluster
falls off with a power law with exponent $\overline{\gamma}$. The
average cluster size~\eqref{timothy2} in the diluted network, on
the other hand, can be further worked out by differentiating
Eqs.~\eqref{H} with respect to $x$:
\begin{align}
\mathcal{S}  =  1 + \frac{f\langle k\rangle}{1 + F_1'(0)}.
\label{avcl}
\end{align}
Hence the average cluster size diverges when
\begin{align}
1 = -F_1'(0).
\end{align}
This is yet another way of writing the Molloy-Reed criterium
Eq.~\eqref{molloyreed}.

\subsection{Full Derivation of Self-consistent Equations}\label{fullderivation}
We prove now that the self-consistent Eqs.~\eqref{H} are only
valid in case no correlations are introduced in the reconstructed
network, or, when $f\rho_{kq}=\rho_k\rho_q$, as is valid for the
simultaneous method only. Here we will give a precise derivation
of the self-consistent Eqs.~\eqref{H}, thereby  taking into
account the degree-dependence of the functions $H_1$.

Let us first look at the generating function
$\hat{H}_1^{q\rightarrow k}(h)$ for the probability that an edge,
which connects nodes of degree $q$ and $k$, branches out in a
cluster of a given edge number along the node of degree $k$. It is
readily derived that $\hat{H}_1^{q\rightarrow k}$ satisfies the
equation:
\begin{align}\label{hoeplala}
\hat{H}_1^{q\rightarrow k}(h)=e^{-h} \left(1+\sum_{\hat{k}=m}^K
P_n(\hat{k})\rho_{k\hat{k}}[\hat{H}_1^{k\rightarrow
\hat{k}}(h)-1]\right)^{k-1}.
\end{align}
We proceed by defining $H_1^q(h)=\sum_{k} P_n(k)
\rho_{qk}\hat{H}_1^{q\rightarrow k}(h)/\rho_q$, such that we
arrive at a set of self-consistent equations for each value of
$q$:
\begin{align}\label{exactH1}
H_1^q(h)&=e^{-h}\sum_{k=m}^K\frac{P_n(k)
\rho_{qk}}{\rho_q}\left(1+\rho_k[H_1^k(h)-1]\right)^{k-1}.
\end{align}
Note that, as derived in Sec.~\ref{characteristics}, no
correlations are induced during depreciation when
$f\rho_{kq}=\rho_k\rho_q$. In that case, this equation reduces to
Eq.~\eqref{H0}. After solving Eq.~\eqref{exactH1} with respect to
$H_1^k$ for all values of $k$, we can also calculate:
\begin{align}
H_0(h)&=\sum_{k=m}^K
e^{-h}P(k)\left[1+\rho_k[H_1^k(h)-1]\right]^{k}.
\end{align}
Again, this expression reduces to Eq.~\eqref{H1} in case of
correlation-free depreciation when $H_1^k$ is independent of $k$.

Below the percolation transition, a trivial solution exists:
$H_1^{k}(0)=1$. This solution, however, turns unstable at the
percolation threshold. The threshold value may be derived by
linearization of $H_1^{k}(0)$ around its equilibrium value:
$H_1^{k}(0)=1-\varepsilon_{k}$ with $\varepsilon_{k}\ll 1$. The
percolation criterion is then:
\begin{align}\label{stelsel}
\varepsilon_{q}&=\sum_{k=m}^K\frac{P_n(k)
\rho_{qk}\rho_{k}}{\rho_{q}}(k-1)\varepsilon_{k}.
\end{align}
Again, in absence of correlations in the network, that is, when
the criterion $f\rho_{qk}=\rho_k\rho_q$ is satisfied, this reduces
to the earlier encountered Molloy-Reed criterion of
Eq.~\eqref{starsturntodust}. Eq.~\eqref{stelsel} is the criterion
for the percolation threshold for correlated systems, such as the
one created using the sequential method. However, solving
Eq.~\eqref{stelsel} to obtain $f_c$ constitutes a rather difficult
task.

\subsection{Original and Depreciated Network}\label{upgrade}

We show now that another approach exists by which one easily
derives the self-consistent equations characterizing the network.
This method is closer to the one followed in other works.

Let us call $R^{i\rightarrow j}$ the probability that an edge in
the network does not lead to a vertex connected via the remaining
edges to the giant component (infinite cluster) and $\rho_{ij}$
the probability that the edge between nodes $i$ and $j$ is active.
Then, following the edge along node $j$, one finds:
\begin{align}
R^{i\rightarrow j}=1-\rho_{ij}+\rho_{ij}\prod_{z=1\ldots k_j-1}
R^{j\rightarrow z}.
\end{align}
This type of equation was already obtained for the $q\rightarrow
1$ limit of the $q$-state Potts model~\cite{dorogovtsev4}. Using
the tree approximation, we can rewrite everything as a function of
the degrees of the nodes:
\begin{align}\label{selfish}
R^{q\rightarrow k}=1-\rho_{qk}+\rho_{qk}\left[\sum_{\hat{k}=m}^K
P_n(\hat{k})R^{k\rightarrow \hat{k}}\right]^{k-1}.
\end{align}
This self-consistent set of equations is equivalent to the ones
that we obtained in Eq.~\eqref{exactH1}. Indeed, after the
transformation $R^{q\rightarrow
k}-1=\rho_{qk}(\hat{R}^{q\rightarrow k}-1)$, we find:
\begin{align}\label{selfish2}
\hat{R}^{q\rightarrow k}=\left[1+\sum_{\hat{k}=m}^K
P_n(\hat{k})\rho_{k \hat{k}}(\hat{R}^{k\rightarrow
\hat{k}}-1)\right]^{k-1},
\end{align}
which is exactly the same as Eq.~\eqref{hoeplala} for $h=0$.

The purpose of this derivation is to show that our self-consistent
equation~\eqref{hoeplala} is in agreement with
Eq.~\eqref{selfish}. For degree-independent $R$ and $\rho$, an
equation similar to Eq.~\eqref{selfish} appears frequently in the
literature. The difference between Eq.~\eqref{hoeplala} and
Eq.~\eqref{selfish2} is that $\hat{H}_1$ is normalized with
respect to the depreciated network, whereas $\hat{R}$ is
normalized with respect to the original network.

\section{Equivalence with the Potts model}\label{kastfort}
There exists an equivalence between edge percolation and the
$q\rightarrow 1$ limit of the $q$-state Potts model. This
connection was first worked out by Fortuin and Kasteleyn in
Ref.~\onlinecite{fortuin}. Although initially used for lattice
models, the connection was very general and is valid for any
network~\cite{fortuin,wu}. Moreover, their proof can easily be
generalized to incorporate edge-dependent coupling constants and
edge-dependent removal into the Potts model and the percolation
model, respectively. We explain here in more detail this
equivalence and reformulate our percolation problem as a spin-like
problem which will allow us to derive critical exponents and
compare them with the ones obtained for the Potts model. We will
also find support for our simple scaling relation using exponent
$\overline{\gamma}$ (Eq.~\eqref{mappy}), as was already
encountered in studies concerning degree-dependent Ising
interactions on scale-free networks \cite{giuraniuc1,giuraniuc2}.
Note that the Ising model~\cite{leone, dorogovtsev3} and the Potts
model~\cite{igloi,dorogovtsev4}, together with their critical
properties, were already studied on scale-free networks.

The Potts model can be seen as a generalization of the Ising model
in which each site $i$ has a spin $\sigma_i$. In the Potts model,
these spins can take $q$ distinct values $0,\dots,q-1$ and the
Potts Hamiltonian is:
\begin{align}
\mathcal{H} =-\sum_{<ij>}J_{ij} \delta_{\sigma_i,\sigma_j}-
hk_BT\sum_i \delta_{\sigma_i,0}.
\end{align}
Here $<ij>$ indicates nearest-neighbor sites $i$ and $j$, $J_{ij}$
is the coupling constant and $\delta$ the Kronecker delta
function. Note that the Ising model corresponds to the $q=2$ Potts
model. The Fortuin-Kasteleyn theorem states now that the free
energy of the $q\rightarrow 1$ limit of the $q$-state Potts model
is the same as the ``free energy'' of the percolating network
where the latter is the generating function of the cluster size
distribution function $n_s$:
\begin{align}
\mathcal{F}(f,h)&=\left\langle \sum_s n_s e^{-hs}\right\rangle.
\end{align}
Here the average is performed over all networks in which the
probability to retain the edge between nodes $i$ and $j$ is
$\rho_{ij}$. The parameter $\rho_{ij}$ in the percolation problem
corresponds in the following way to parameters of the Potts
model~\cite{herrmann}~\footnote{This relation can be interpreted
as follows: Consider that the average rate of spin-flipping
between $i$ and $j$ is $J_{ij}/k_BT$, then, in the continuum
limit, the probability $\rho_{ij}$ that the spin will be flipped
after one time unit is equal to
$$
\rho_{ij}=1-\lim_{\delta t\rightarrow 0}(1-J_{ij}\delta
t/k_BT)^{1/\delta t}=1-e^{-J_{ij}/k_BT}.$$}:
\begin{align*}
\rho_{ij}&\quad\leftrightarrow\quad 1 - e^{-J_{ij}/k_BT}.
\end{align*}
From this relation, we can immediately identify the probability
$\mathcal{P}_{\infty}$ for a node to be in the infinite cluster
and the average cluster size $\mathcal{S}$, earlier introduced in
Eq.~\eqref{richard}. As we are interested in the behavior near
criticality, we introduce
\begin{align}
\epsilon=f-f_c,
\end{align}
and obtain
\begin{subequations}
\begin{align}
\mathcal{P}_{\infty}(\epsilon)&=1+\left.\frac{\partial
\mathcal{F}}{\partial
h}\right|_{h=0},\\
\mathcal{S}(\epsilon)& =\left.\frac{\partial^2
\mathcal{F}}{\partial h^2}\right|_{h=0}.
\end{align}
\end{subequations}
Note also that $\mathcal{F}(\epsilon,0)$ gives the total number of
finite clusters.

\section{Scaling Theory and Critical Exponents}\label{critexponents}
In the following section, we introduce finite-size scaling in
order to find critical exponents near the percolation transition.
In order to solve the scaling relation, we use a Landau-like
theory which we derive from the exact relations~\eqref{H}. We
follow closely the approach presented in
Refs.~\onlinecite{botet1,botet2} and~\onlinecite{hong} for
finite-size scaling in systems with dimensions above the upper
critical dimension. However, we will find that the forms of the
Landau-like theories of Refs.~\onlinecite{hong,igloi}
and~\onlinecite{goltsev2} were too limited for studying the
percolation transition in case the distribution function has a
very fat tail, that is when $2<\gamma<3$.

According to finite-size scaling, the free energy $\mathcal{F}$ of
a large but finite network with $N$ nodes close to criticality can
be written in the general form~\cite{stauffer}:
\begin{align}\label{freeenergy}
\mathcal{F}(\epsilon,h)=N^{-1}\mathbb{F}\left(\epsilon
N^{1/\nu_{\epsilon}},hN^{1/\nu_{h}}\right),
\end{align}
where $\mathbb{F}$ is a well-behaved function. The variable
$\epsilon N^{1/\nu_{\epsilon}}$ originates from the existence of a
``correlation number'' $N_{\xi}$ (instead of a correlation length)
which scales as $N_{\xi}\propto \epsilon^{-\nu_{\epsilon}}$ such
that the first variable of $\mathbb{F}$ can be rewritten as
$(N/N_{\xi})^{1/\nu_{\epsilon}}$~\cite{botet1,botet2}. It is then
obvious that close to criticality, the (singular part of the) free
energy scales as:
\begin{subequations}\label{robert}
\begin{align}
\mathcal{F}(\epsilon,0)&\propto\epsilon^{\nu_{\epsilon}},\\
\mathcal{F}(0,h)&\propto h^{\nu_{h}}.
\end{align}
\end{subequations}
As a second scaling ansatz, we assume that, in the macroscopic
limit, the scaling of the cluster size distribution $n_s$ can be
written as~\footnote{Note that for $d$-dimensional lattices with
$d>8$, the cluster size distribution can indeed be written in this
form~\cite{stauffer,harris}. Since our considered random networks
have essentially an infinite dimension, this scaling ansatz is
justified.}:
\begin{align}\label{clusterdistribution}
n_s(\epsilon)=s^{-\tau}\mathbb{G}(\epsilon s^{\sigma}),
\end{align}
where again $\mathbb{G}$ is a well-behaved function. The form of
$n_s$ one usually has in mind is $n_s \propto
s^{-\tau}e^{-s/s^*}$~\cite{newman}, which is essentially a damped
power-law with cutoff $s^*$, valid for large cluster sizes. At
criticality very large clusters arise, caused by the diverging
cutoff $s^*$ according to $s^*\propto \epsilon^{-1/\sigma}$ such
that $n_s(0)\sim s^{-\tau}$.

Using the analogy with the Potts model, we define the usual
critical exponents $\alpha$, $\gamma_p$, $\beta$ and $\delta$ for
the percolation problem as:
\begin{subequations}
\begin{align}
&\mathcal{F}(\epsilon,0)\sim
\epsilon^{2-\alpha},\\
&\mathcal{P}_{\infty}(\epsilon)\sim
\epsilon^{\beta},\\
&\mathcal{S}(\epsilon)\sim
\epsilon^{-\gamma_p},\\
&\left.\frac{\partial \mathcal{F}}{\partial
h}\right|_{\epsilon=0}+1 \sim h^{1/\delta}.
\end{align}
\end{subequations}
Using the scaling forms of Eqs.~\eqref{freeenergy}
and~\eqref{clusterdistribution}, standard techniques provide us
with exponent relations by which all critical exponents can be
related to $\nu_{h}$ and $\nu_{\epsilon}$. One arrives
at~\cite{stauffer}
\begin{subequations}\label{exponents}
\begin{align}
\beta &=\nu_{\epsilon}(1-\nu_{h}^{-1}),\\
\gamma_p&=\nu_{\epsilon}(2\nu_{h}^{-1}-1),\\
\alpha &=2-2\beta-\gamma_p,\\
\sigma &=(\beta+\gamma_p)^{-1},\\
\tau &=2+\beta(\beta+\gamma_p)^{-1},\\
\delta &=(\beta+\gamma_p)/\beta,\\
\nu_f &=\beta+\gamma_p.
\end{align}
\end{subequations}
The last exponent $\nu_f$ is related to the usual fractal
dimension $d_f$ in the same way that $\nu_{\epsilon}$ is related
to the usual dimension $d$ and quantifies how the cluster size $s$
scales with $\epsilon$ close to criticality, i.e., $s \propto
\epsilon^{-\nu_f}$. Note that the cutoff size $s^*$ scales like
$s$, since $\nu_f = 1/\sigma$.

The problem we are left with now is to find the scaling exponents
$\nu_{h}$ and $\nu_{\epsilon}$ for percolation on scale-free
networks. This can be done in an exact way since we know the
equation of state from Eq.~\eqref{H} as a function of the order
parameter:
\begin{align}
\psi(\epsilon,h)=1-H_1(\epsilon,h).
\end{align}
As we are merely interested in the behavior near the transition
where $\epsilon\ll 1$, $h\ll 1$ and $\psi\ll 1$, we can expand
Eq.~\eqref{H0}. For the case $\overline{\gamma}>3$, we find the
form~\cite{cohen3}:
\begin{align}\label{eos}
h =-c_1\epsilon \psi +c_2\psi^2+ ... +
c_s\psi^{\overline{\gamma}-2}+\ldots,
\end{align}
in which all $c_i$ as well as the coefficient of the singular
term, $c_s$, are positive constants. This equation also follows
from minimization of the free energy:
\begin{align}\label{tristan}
\mathcal{F}(\epsilon,h)\propto -h\psi-\overline{c}_1\epsilon
\psi^2 +\overline{c}_2\psi^3+...
+\overline{c}_s\psi^{\overline{\gamma}-1}+\ldots,
\end{align}
with respect to the order parameter $\psi$.

We distinguish two cases now. First, when $4<\overline{\gamma}$,
we know that $f_c$ is finite and the relevant part of the equation
of state for $\psi$ becomes:
\begin{align}
h=-c_1\epsilon \psi +c_2\psi^2.
\end{align}
Solving for $\psi$, and substitution into Eq.~\eqref{tristan} one
then simply finds that the free energy scales as:
\begin{subequations}\label{robert}
\begin{align}
\mathcal{F}(\epsilon,0)&\propto \epsilon^{3},\\
\mathcal{F}(0,h)&\propto h^{3/2}.
\end{align}
\end{subequations}
In other words, when $4<\overline{\gamma}$, we find that
$\nu_{\epsilon}=3$ and $\nu_{h}=3/2$. From these two exponents,
and using Eq.~\eqref{exponents}, we list all other exponents in
the last column of Table $1$. Note that, as expected, these
exponents agree with the usual mean-field results for
percolation~\cite{stauffer,essam}.

Secondly, when $3<\overline{\gamma}<4$, the relevant part of
Eq.~\eqref{eos} reduces to
\begin{align}
h =-c_1\epsilon \psi +c_s\psi^{\overline{\gamma}-2},
\end{align}
from which follows that
\begin{subequations}\label{robert2}
\begin{align}
\mathcal{F}(\epsilon,0)&\sim \epsilon^{\frac{\overline{\gamma}-1}{\overline{\gamma}-3}},\\
\mathcal{F}(0,h)&\sim
h^{\frac{\overline{\gamma}-1}{\overline{\gamma}-2}}.
\end{align}
\end{subequations}
Therefore, we come to:
\begin{align}
\nu_{\epsilon}=\frac{\overline{\gamma}-1}{\overline{\gamma}-3}\text{
and  } \nu_{h}=\frac{\overline{\gamma}-1}{\overline{\gamma}-2},
\end{align}
and we obtain all other exponents as given in the second column of
Table $1$.

\renewcommand{\arraystretch}{2.5}

\begin{table}[ht]
\caption{Critical Exponents\label{table1}} \centering
\begin{tabular}{l ||  c | c | c}
\hline\hline

& $2<\overline{\gamma}<3$

& $3<\overline{\gamma}<4$

& $\overline{\gamma}>4$ \\

\hline

$\beta$

& $\dfrac{1}{3-\overline{\gamma}}$

& $\dfrac{1}{\overline{\gamma}-3}$

& $1$

\\

\hline

$\tau$

& $\dfrac{2\overline{\gamma}-3}{\overline{\gamma}-2}$

& $\dfrac{2\overline{\gamma}-3}{\overline{\gamma}-2}$

& $5/2$

\\

\hline

$\sigma$

& $\dfrac{3-\overline{\gamma}}{\overline{\gamma}-2}$

& $\dfrac{\overline{\gamma}-3}{\overline{\gamma}-2}$

& $1/2$

\\
\hline

$\alpha$

& $-\dfrac{3\overline{\gamma}-7}{3-\overline{\gamma}}$

& $-\dfrac{5-\overline{\gamma}}{\overline{\gamma}-3}$

& $-1$

\\
\hline

$\gamma_p$

& $-1$

& $1$

& $1$

\\
\hline

$\delta$

& $\overline{\gamma}-2$

& $\overline{\gamma}-2$

& $2$

\\
\hline

$\nu_f$

& $\dfrac{\overline{\gamma}-2}{3-\overline{\gamma}}$

& $\dfrac{\overline{\gamma}-2}{\overline{\gamma}-3}$

& $2$

\\

\end{tabular}
\label{table:nonlin} % is used to refer this table in the text
\end{table}

Lastly, in case $2<\overline{\gamma}<3$, the critical fraction
$f_c=0$. Again, we expand Eq.~\eqref{H0} using the small
parameters $\epsilon\ll 1$, $|h|\ll 1$ and $\psi\ll 1$. The
equation of state for $\psi$ and the associated free
energy~\eqref{tristan} become:
\begin{subequations}
\begin{align}
h =&-c_s(\epsilon\psi)^{\overline{\gamma}-2}+\psi+\ldots,\\
\mathcal{F}(\epsilon,h)\propto
&\,\epsilon\left(-h\psi+\overline{c}_s\epsilon^{\overline{\gamma}-2}\psi^{\overline{\gamma}-1}+\psi^2/2\right)+\ldots,
\end{align}
\end{subequations}
where $c_s>0$. It follows that
\begin{subequations}\label{robert2}
\begin{align}
\mathcal{F}(\epsilon,0)&\sim \epsilon^{\frac{\overline{\gamma}-1}{3-\overline{\gamma}}},\\
\left.\mathcal{F}(\epsilon,h)\right|_{\epsilon\rightarrow 0}&\sim
|h|^{\frac{\overline{\gamma}-1}{\overline{\gamma}-2}}.
\end{align}
\end{subequations}
In the last expression, the limit $\epsilon\rightarrow 0$ is only
taken in the free energy and $h$ is taken small and negative such
that $\psi$ is still positive. Therefore,
\begin{align}\label{nuuus}
\nu_{\epsilon}=\frac{\overline{\gamma}-1}{3-\overline{\gamma}}\text{
and  } \nu_{h}=\frac{\overline{\gamma}-1}{\overline{\gamma}-2}.
\end{align}
The other exponents are listed in the first column of Table $1$.
It must be noted here that in practice, the exponents of
Eq.~\eqref{nuuus} may be impossible to find with the
configurational model in case we start from a robust network
($2<\gamma<3$). This stems from the fact that a structural cutoff
for the maximally allowed degree $K$ must be introduced to obtain
an uncorrelated network. Such cutoff can be of the form $K\sim
N^{1/\omega}$  with $\omega\in [2,\infty[$. However, it is
well-known that the cutoff affects the critical
exponents~\cite{castellano}. Indeed, performing the averages in
Eq.~\eqref{criterion584} with use of the cutoff, one readily
obtains $f_c\propto N^{(1-\alpha)(\overline{\gamma}-3)/\omega}$
(cf.~Eq.~\eqref{fcN}) and therefore:
\begin{align}\label{nuu}
\nu_{\epsilon}=-\frac{\omega}{(1-\alpha)(\overline{\gamma}-3)}.
\end{align}
In case we start from a fragile network ($\gamma>3$), $\omega$
equals $\gamma-1$ and the $\nu_{\epsilon}$ of Eq.~\eqref{nuuus} is
retrieved. Therefore, degree-dependent edge removal may be used as
a tool to observe critical exponents in the delicate regime
$2<\overline{\gamma}<3$, \textit{without} the use of a structural
cut-off~\cite{castellano}. In our simulations as presented in
Sect.~\ref{comparison}, we take $\omega=2$ in case $\gamma<3$.

The exponents obtained for the case $\overline{\gamma}>3$ reduce
to the ones for \textit{node} percolation in the limit $\alpha =
0$ when $\overline{\gamma}=\gamma$~\cite{cohen3}. However, in the
limit $\alpha=0$ the exponents in the first column do not coincide
with those given in Refs.~\onlinecite{lee}
and~\onlinecite{cohen3}.

%~\footnote{Although the exponent $\beta$ seems to agree with the
%exponent found by Cohen et al.~\cite{cohen3} for random {\em node}
%percolation, their result appears to be in error. The erroneous
%exponent results when $P_{\infty} \propto \epsilon$ is used
%instead of the appropriate expression $P_{\infty} \propto
%\delta\epsilon$, which would lead to the {\em modified} critical
%exponent $\beta =(4-\overline{\gamma})/(3-\overline{\gamma})$.
%This correct exponent is also by found in
%Ref.~\onlinecite{goltsev}.}.

Close to the percolation transition, it is possible, with the use
of Eqs.~\eqref{richard} to calculate the average degree in the
giant cluster. We can identify this as
\begin{align}
\lim_{f\rightarrow
f_c}\frac{fN_e(1-(H_1)^2)}{N(1-H_0)}=\frac{(f_c+\epsilon)\langle
k\rangle 2N }{N\langle k\rangle (f_c+\epsilon)}=2.
\end{align}
This however, must not be confused with the Molloy-Reed Criterion,
which states that the average degree of a \textit{neighboring site
in the entire network} has degree two.

In sum, we have now calculated the most important critical
exponents for a percolation process. Extra support for our
critical exponents comes from scaling relations and the connection
with the Potts model.

There is one feature which appears in all the calculated
exponents: the only dependence on $\alpha$ arises through the
exponent $\overline{\gamma}$. Random percolation on a network with
degree exponent $\overline{\gamma}$ gives the same critical
exponents as percolation with bias exponent $\alpha$ on a network
with degree exponent $\gamma$. This equivalence was found before
for degree-dependent interactions on scale-free
networks~\cite{giuraniuc1,giuraniuc2}. It is not surprising that
the same behavior appears both for edge percolation and for
degree-dependent interactions, since both can be linked with the
Fortuin-Kasteleyn construction.

\section{Comparison with Numerical Results}\label{comparison}
\begin{figure}
 \begin{center}
  \includegraphics[width = .4 \textwidth]{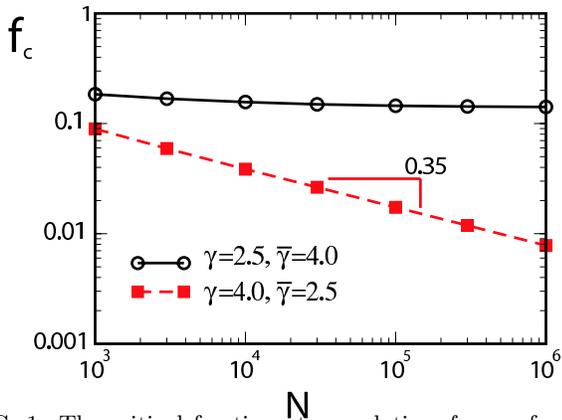}
\caption{The critical fraction at percolation $f_c$ as a function
of the network size $N$. To compute this critical fraction we
average over $10^4$ network realizations for each set of
parameters and apply for each network the percolation process 100
times. No cutoff was introduced for the maximal node degree and
network reconstruction was done with the sequential
approach.\label{pcN}}
 \end{center}
\end{figure}

In this section we test the previously derived analytical results
using simulations. The networks are generated using the
uncorrelated configurational model which was introduced in
Sect.~\ref{sec_intro}. Each simulation involves three free
parameters: the degree exponent $\gamma$ of the network, the
minimal node degree $m$ and the number of vertices $N$. Unless
mentioned otherwise, we set $m=1$. In the configuration model, the
degrees of the nodes are determined initially from the discrete
degree distribution~\footnote{Note that it is essential to take
the distribution function to be defined on a {\em discrete}
support ($k= 0,1,2, ...$) in order to match the simulations with
our analytical work.} and then the connections are assigned at
random. To obtain an uncorrelated network, the maximal degree is
set to $\sqrt{N}$ when $2<\gamma<3$~\cite{catanzaro}. In some
simulations (see Figs.~\ref{pcN} and~\ref{cross-overfig}) no
degree cutoff was imposed. If $\gamma \geq 3$, the maximal degree
is simply $N-1$. Both the sequential and the simultaneous approach
are implemented.

\subsection{Sequential Approach}
In this first subsection, we discuss simulation results concerning
the sequential approach. Most of these results can also be found
in Ref.~\onlinecite{moreira}.

\subsubsection{Scaling of the Critical Point}

First, we investigate the finite-size behavior of the critical
fraction $f_c$  of nodes. The results for the sequential approach
are shown in Fig.~\ref{pcN}. For networks with $\gamma=2.5$
submitted to CB with an effective value $\overline{\gamma}=4$
(continuous black line), we observe that the critical fraction
$f_c$ converges to a finite value as $N$ grows, confirming the
conjecture that a robust network may turn fragile under CB. In the
opposite case, a network with $\gamma=4$ submitted to PB with an
effective $\overline{\gamma}=2.5$ (dashed red line), has a
critical fraction that decays with the vertex number $N$ as a
power-law, $f_c\sim{N^{-1/\nu_{\epsilon}}}$. The best fit to the
data in this case results in $1/\nu_{\epsilon}={0.35\pm{0.02}}$,
consistent with the value $1/3$ expected from Eq.~\eqref{fcN}.
This result shows that a fragile network under PB will behave in
the same fashion as a robust network with a degree distribution
controlled by $\overline{\gamma}$ under random failure
\footnote{We remark that the choice of $\alpha$ in the PB case
here is a little bit special, since $\alpha = 3-\gamma$, which
means that this case is just the borderline case between the
strong PB and the weak PB discussed in Ref.~\onlinecite{moreira}.
The simulations indicate that this borderline case belongs to the
strong PB regime.}. Note that this simulation result confirms
Eq.~\eqref{fcN}, although this equation was derived for the
simultaneous approach. Indeed, to deduce Eq.~\eqref{fcN},
correlations in the diluted network are neglected. These
simulation results indicate that this is an acceptable
approximation.

\subsubsection{Properties of the Diluted Network}
\begin{figure}[t]
 \begin{center}
  \includegraphics[width = .4 \textwidth]{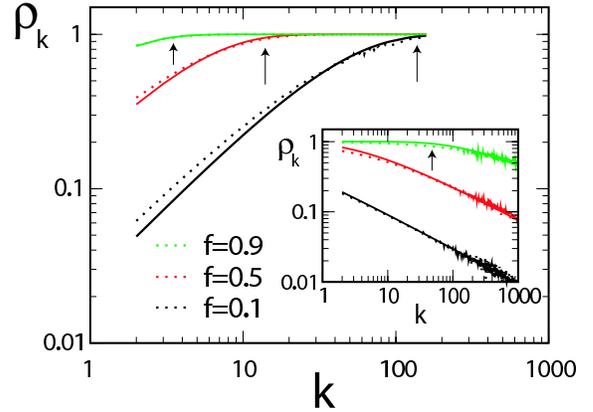}
\caption{The probability $\rho_k$ to retain a node after
depreciation (see Eq.~\eqref{marginal}) as a function of its
original degree $k$ using the sequential approach. The main panel
shows results for networks with $\gamma=4$ submitted to PB with
$\alpha=-1$. The dots indicate the simulation results for ten
network realizations and ten percolation routines. The continuous
lines are the best fit to the data of Eq.~\eqref{ghinzu}. The
inset shows the same but for a network with $\gamma=2.5$ subjected
to CB with $\alpha=0.5$. In both cases, $m = 2$ and $N= 10^5$. The
arrows indicate the crossover value $k_{\times}$. No cutoff was
introduced for the maximal node degree. \label{cross-overfig}}
 \end{center}
\end{figure}

During our theoretical discussion of the sequential approach,
certain properties of the diluted network became apparent, such as
a cross-over behavior as a function of the degree $k$ for the mean
edge preservation probability $\rho_k$. The probability $\rho_k$
can easily be inferred from our simulations by calculating the
ratio of the new node degree $\overline{k}$ to the old node degree
$k$ for each node and averaging over all nodes with the same
degree. The result is shown in Fig.~\ref{cross-overfig} in which
the expected crossover behaviors are marked by arrows and the
continuous lines are fits to the data of Eq.~\eqref{ghinzu}. When
CB is applied (inset), a cross-over between a regime with
$\rho_k\approx 1$ to a decreasing power law is found.

\begin{figure}[t]
\begin{center}
 \includegraphics[width = .35 \textwidth]{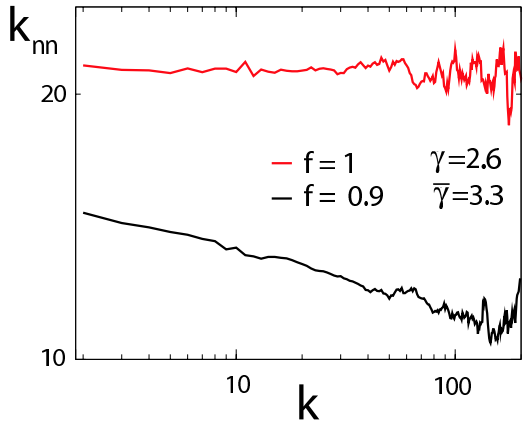}
\caption{Mean nearest-neighbor degree $k_{nn}$ as a function of
the degree of a node $k$, both for the original network (red line)
and a network where only a fraction $f = 0.9$ of the links is
present (black line). Clearly correlations which give rise to
disassortative mixing emerge in the depreciated network. We used
an original network constructed with the uncorrelated
configuration model with $\gamma = 2.6$, $m=2$ and $N = 10^5$
which is diluted using sequential biased percolation with $\alpha
= 0.3$, such that $\overline{\gamma} = 3.3$. The result was
obtained with two network realizations, on both of them the
percolation process was applied four times. To reduce the noise
level, the mean-neighbor degree is averaged over eight successive
values.\label{correl4}}.
\end{center}\end{figure}

As a second characteristic of diluted networks, the emergence of
correlations in the diluted network is discussed. The theoretical
result of Eq.~\eqref{disassort} suggests disassortative mixing in
the diluted network in case central bias is applied. To observe
these correlations in the simulations, the mean nearest-neighbor
degree is calculated as a function of the node degree. The result
of our simulation is shown in Fig.~\ref{correl4}. As expected, no
correlations are present in the original network (top curve in
red, $f=1$). However, with 10\% of its edges removed, the mean
nearest-neighbor degree in the diluted net clearly decreases as
the degree of the node increases (bottom curve in black). This is
a clear indication of disassortative mixing after sequentially
removing a certain fraction of edges and this confirms our
theoretical prediction.

\subsection{Simultaneous Approach}
\subsubsection{Comparison with Theory and Sequential Approach}
\begin{figure}
\subfigure{\includegraphics[ width = .4 \textwidth]{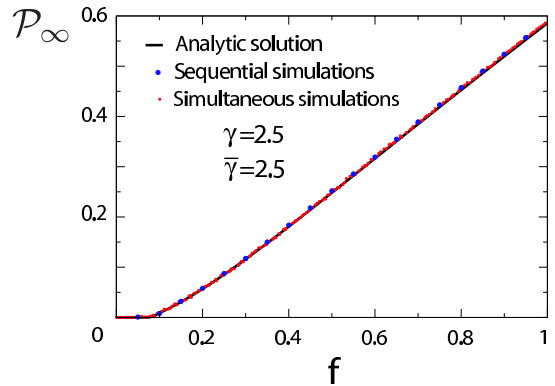}}

\caption{The probability that a node belongs to the giant cluster,
$\mathcal{P}_{\infty}$, as a function of the fraction of retained
links $f$ for unbiased  percolation ($\alpha=0$). We compare the
sequential approach simulations (blue dots), the iterative
simultaneous approach simulations (red dots) and the theory of
generating functions (black line). We used an original network
with $\gamma = 2.5$ and $N = 10^5$. The analytical results up to
the fraction $f_u=1$ are obtained with the generating functions
theory. All results are in very good mutual agreement.}
\label{orderparam1}
\end{figure}

\begin{figure}
\subfigure{\includegraphics[width = .4
\textwidth]{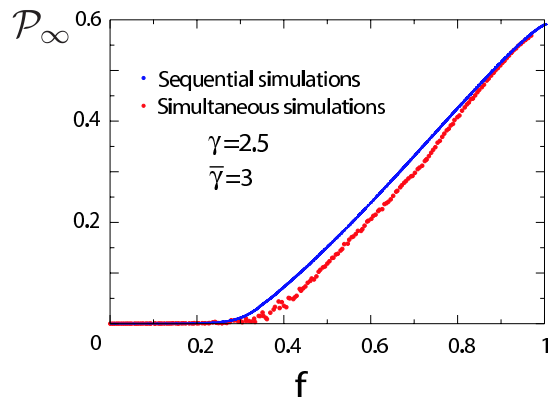}}\caption{The probability that a node belongs to
the giant cluster, $\mathcal{P}_{\infty}$, as a function of the
fraction of retained links $f$ for biased percolation with $\alpha
= 0.25$. We compare the sequential (blue line) and simultaneous
(red dots) approach simulations. We used an original network with
$\gamma = 2.5$ and $N = 10^5$.}\label{orderparam2}
\end{figure}

\begin{figure}
\subfigure{\includegraphics[width = .4 \textwidth]{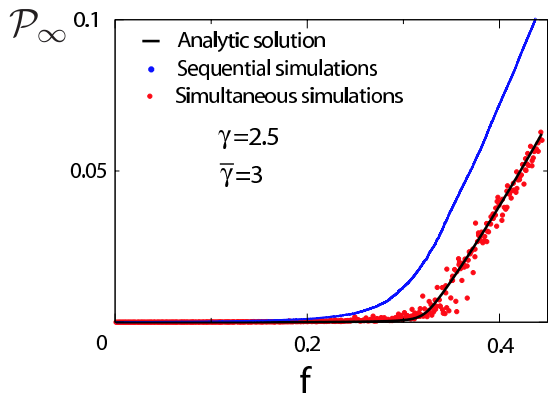}}
\caption{The probability that a node belongs to the giant cluster,
$\mathcal{P}_{\infty}$, as a function of the fraction of retained
links $f$ for biased percolation with $\alpha = 0.25$. We compare
the sequential approach simulations (blue line), the iterative
simultaneous approach simulations (red dots) and the theory of
generating functions (black line). We used an original network
with $\gamma = 2.5$ and $N = 10^5$. The analytical results up to
the fraction $f_u=0.44$ are obtained with the generating functions
theory. This figure presents more detail of the critical region of
Fig.~\ref{orderparam2}.}\label{orderparam3}
\end{figure}

This subsection deals with the iterated simultaneous approach as
introduced at the end of Sect.~\ref{simultaneousapproach}. To
examine the percolation transition, we search for the probability
to belong to the largest cluster, $\mathcal{P}_{\infty}$, as a
function of the fraction $f$ of included edges. Results are given
in Figs.~\ref{orderparam1},~\ref{orderparam2}
and~\ref{orderparam3}.

For random edge removal it is shown in Fig.~\ref{orderparam1} that
the simultaneous approach coincides with the sequential approach;
in that case, only one iteration is necessary to attain $f=1$.
When, on the other hand, $\alpha>0$, the simultaneous approach can
only be used up to a certain value of $f_u$, smaller than one (See
Eq.~\eqref{fff}). However, the iterative procedure can be used
until all links are included. In general, we can include over 98\%
of the links with a relatively small number of
weight-recalculations or iterations. For instance, in case a
network with $\gamma = 2.5$ is subjected to percolation with
$\alpha = 0.2$, one finds that $f_u = 0.51$ and 15 iterative steps
are necessary to reach $f = 0.98$.

The results of such iterative simultaneous approach can be found
in figure \ref{orderparam2}. It is immediately clear that the
sequential and the simultaneous no longer coincide. This is not
surprising since the definitions of edge retaining probabilities
$\rho_{ij}$ are different for the sequential and the simultaneous
approaches when $\alpha \neq 0$. Furthermore, the sequential
approach introduces correlations which are absent in the
simultaneous approach. Note, however, that the difference between
the two approaches is not a mere consequence of the appearance of
correlations in the sequential approach. Indeed, if only the
(dissassortative) correlations were present, the sequential
approach should have a larger critical fraction $f_c$ than the
simultaneous approach~\cite{goltsev}; yet, we find the inverse to
be true as evidenced in Fig.~\ref{orderparam3}. At the point at
which the weights are recalculated, the giant cluster probability
$\mathcal{P}_{\infty}$ of the iterative simultaneous approach
shows kinks as a function of $f$. For instance, a conspicuous kink
appears at $f_u = 0.69$ in Fig.~\ref{orderparam3}. The iterative
simultaneous and sequential methods approach each other as
$f\rightarrow 1$ and coincide at $f=1$.

Note that the critical fraction $f_c$ which can be extracted from
Figs.~\ref{orderparam1},~\ref{orderparam2} and~\ref{orderparam3}
is nonzero although it is expected to be zero for $2\leq
\overline{\gamma}\leq 3$. This is a consequence of the finite-size
effects which cause $f_c$ to scale with the system size according
to Eq.~\eqref{fcN} (see also Fig.~\ref{pcN}).

Although there are differences between the sequential and the
simultaneous approach, the differences are clearly not very large.
The largest \textit{relative} deviations occur around $f_c$ while
the largest \textit{absolute} deviations are situated around the
lowest $f_u$ and are typically not more than 10\%. Although we
find the critical fraction $f_c$ of the simultaneous approach to
be always larger than the one of the sequential approach, both
values deviate by less than 10\%. We conclude that the
simultaneous and the sequential approach do differ, but the
differences are not large and both approaches are qualitatively
similar.

Analytical results for the probability of the largest cluster
$\mathcal{P}_{\infty}$ in the regime $f<f_u$ can be obtained by
solving Eq.~\eqref{H0} numerically for $H_1(1)$ which is then
introduced in Eq.~\eqref{H1}. For random edge removal
($\alpha=0$), $f_u = 1$ and thus a theoretical result is available
for all $f$-values. Moreover, also for other values of $\alpha$
the generating functions theory calculation (see black line in
Fig.~\ref{orderparam3}) agrees well with simulation results
throughout the entire reconstruction process. The theoretical
model is thus justified by the simulations.

\begin{figure}
 \begin{center}
  \includegraphics[width = .4 \textwidth]{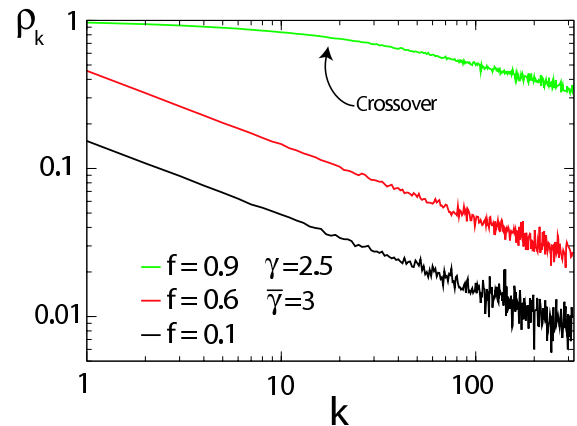} \caption{The probability $\rho_k$ to retain a node after
depreciation (see Eq.~\eqref{marginal}) as a function of its
original degree $k$ for different values of $f$ and using the
iterated simultaneous approach. We used a network with $\gamma =
2.5$, $\alpha=0.5$ and $N = 10^5$ and averaged over ten network
realizations. Results are averages over ten percolation
simulations. Since $f_u = 0.44$, the first regime can be reached
in a single sweep and thus no cross-over emerges for $f=0.1$. The
cross-over (indicated by arrow) becomes apparent for $f=0.9$ when
ten iterative steps are performed. This figure should be compared
with the inset of Fig.~\ref{cross-overfig} where the sequential
approach was used.\label{crossoverpar}}
 \end{center}
\end{figure}

\subsubsection{Properties of the Diluted Network}
We end the section with an overview of the properties of the
diluted network. The focus lies again on the appearance of
correlations and on the cross-over. In the simultaneous approach
(with $f<f_u$), no cross-over can appear for the node retaining
probability $\rho_k$. Indeed, expression \eqref{marginalpaparel}
is exact and predicts a decreasing power law for $\rho_k$, that is
when $f<f_u$. However, this expression is only valid as long as no
recalculation of the weights is performed. As soon as we iterate
the simultaneous approach, the results for the simultaneous and
sequential methods start approaching each other. Since the
sequential approach contains a cross-over, we expect the
appearance of a cross-over in the iterative simultaneous approach.
Analoguous arguments apply to the appearance of correlations in
the diluted network.

\begin{figure}
 \begin{center}
\includegraphics[width =.45\textwidth]{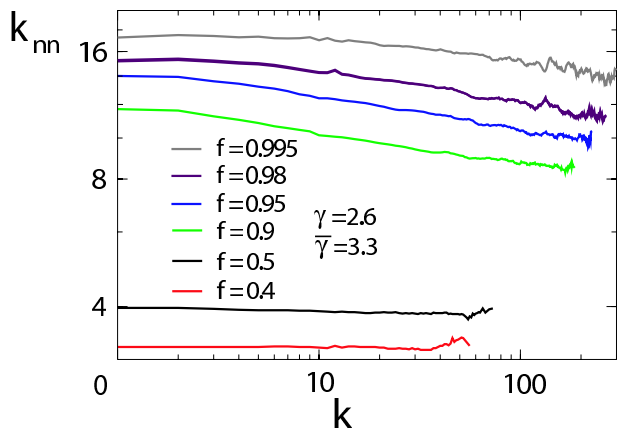}
\caption{Emergence of correlations in the diluted network in the
\textit{iterative} simultaneous approach. This figure shows the
mean nearest-neighbor degree in the diluted network. We used an
original network constructed with the uncorrelated configuration
model with $\gamma = 2.6$, $m=2$ and $N = 10^5$ which is diluted
using the iterative simultaneous approach to biased percolation
with $\alpha = 0.3045$, such that $\bar{\gamma} = 3.3$. The
diluted networks contain, from bottom to top, 40 (red), 50
(black), 90 (green), 95 (blue), 98 (purple) and 99.5 (grey)
percent of the edges of the original network. To obtain such a
network, resp. 1, 2, 5, 8 and 12 iterations have been made.  The
result was obtained with 10 network realizations, on all of them
the percolation process was applied 10 times. To reduce the noise
level, the mean nearest-neighbor degree was averaged over 8
successive values. \label{correl5}}
 \end{center}
\end{figure}

The emergence of a cross-over in the iterative simultaneous
approach is indeed found and shown in Fig.~\ref{crossoverpar}.
When 10\% of the edges are included, no cross-over at all appears.
This is consistent with our arguments since we can simply include
this fraction of edges in one sweep. Also in the second sweep, no
cross-over appears. However, after 10 recalculations of the
weights, the cross-over is undoubtedly present. Once again, our
theoretical arguments are verified. Furthermore, the appearance of
correlations is confirmed by our simulation results as evidenced
in Fig.~\ref{correl5}. Indeed, disassortative correlations are
apparent only for large values of $f$, after several iterations
have been performed. Note also that these correlations disappear
only very slowly upon approach of the point $f=1$ where no
correlations are present (see Fig.~\ref{correl4}).

\section{Conclusions}\label{conclusions}
We have performed a detailed study of biased percolation on
scale-free networks with degree exponent $\gamma$ (with $\gamma
> 2$) and shown that it is possible to tune a robust network fragile and
vice versa. Biased percolation involves degree-dependent removal
of edges, more specifically, we assumed that the probability to
retain an edge is proportional to $(k_ik_j)^{-\alpha}$ with $k_i$
and $k_j$ the degrees of the attached nodes. For $\alpha >0$ the
bias is {\em central} since links between highly connected nodes
are preferentially depreciated, while the converse, {\em
peripheral bias}, corresponds to $\alpha < 0$. Our most important
result is that, at percolation, the properties of a network with
bias exponent $\alpha$ and degree exponent $\gamma$ are the same
as those of a network with bias exponent zero and degree exponent
$\overline{\gamma}=(\gamma-\alpha)/(1-\alpha)$, {\em or} degree
exponent $\gamma$, depending on the sign and the range of
$\alpha$. Let us first elaborate on this main result, in the light
of the present work and recapitulating arguments presented in the
preliminary report \cite{moreira}.

For $\alpha >0$ (with the restriction $\alpha < 1$) the new degree
exponent $\overline{\gamma}
> \gamma$ governs the critical properties of the network that
results when the percolation threshold is reached after biased
depreciation. The exponent $\overline{\gamma}$ controls the
large-degree behavior of the new degree distribution. This new
$\overline P (\overline k)$ is not simply scale-free but {\em
asymptotically} scale-free. There is a cross-over value
$k_{\times}$, so that for $\overline{k} < k_{\times}$ the exponent
$\gamma$ is dominant and for $ \overline{k}
> k_{\times}$ the exponent $\overline{\gamma}$ takes over. For
$\overline{\gamma}
> 3$ the biased depreciation process will reach the percolation threshold
at a finite fraction of retained edges. The network is then
fragile under central bias, regardless of whether the network is
fragile ($\gamma >3$) or robust ($\gamma < 3$) under random
removal. For $\overline{\gamma} < 3$ the biased depreciation will
(in an infinite system) not reach a percolation point since the
critical fraction of retained edges, $f_c$, is zero. The network,
which is robust for random removal, remains robust under centrally
biased removal. However, for a finite system $f_c$ is small but
finite and scales with system size in a manner governed by the
exponent $\overline{\gamma}$, whereas the scaling properties of
$f_c$ for \textit{random} removal are governed by $\gamma <
\overline \gamma$. We have shown, by analytic proof, that
$\overline\gamma$ governs the percolation critical behavior for
the case $\alpha
>0$. Also our numerical results support this conclusion.

For $\alpha < 0$ (with the restriction $2-\gamma < \alpha$), the
critical behavior at percolation is more subtle. Peripherally
biased removal is less destructive than random depreciation and it
is possible that a network that is fragile under random failure
becomes robust when peripheral bias is applied. Noting that
$\overline\gamma < \gamma$ it is obvious that robustness is
preserved for networks with $\gamma < 3$. Conversely, fragility
persists for sure when $3 < \overline\gamma$. However, it is not
obvious what to expect when $\overline\gamma < 3 < \gamma$. The
behavior of the new degree distribution $\overline P (\overline
k)$ for $\overline{k} > k_{\times}$ is, for $\alpha < 0$,
controlled by the exponent $\gamma$, so it would seem that the
properties of the network under random failure are simply not
affected by peripheral bias. However, a finite-size scaling
analysis at criticality reveals that the cross-over value
$k_{\times}$ is larger than the maximal degree in the network,
implying that the new degree distribution will be controlled by
$\overline\gamma$ instead of $\gamma $, {\em provided} $2-\gamma <
\alpha < 3 - \gamma$ (we assume $\gamma > 3$ since this discussion
only makes sense for networks fragile under random failure). We
conclude that sufficiently strong peripheral bias can turn a
fragile network robust, and numerical evidence supports this
conclusion. On the other hand, for $3 - \gamma < \alpha < 0$, it
is not clear whether the network stays fragile under peripherally
biased failure when $\overline \gamma $ drops below 3, which
happens for $\alpha < (3-\gamma)/2$. This problem is still largely
open to future investigation.

Two distinct approaches by means of which a network can be
reconstructed in a degree-dependent manner, the sequential and the
simultaneous approach, have been introduced to perform the edge
removal process. For the sequential approach, we obtained a very
useful {\em analytic approximation} to the marginal distribution
$\rho_k$, which is the mean probability that an edge connected to
a node with degree $k$ is present in the network after
reconstruction. This analytic form clearly features the cross-over
value $k_{\times}$ which plays a crucial role in the network
properties. The simultaneous approach, which is a simpler scheme
useful for $\alpha
> 0$ and for edge number fractions below a value dependent on
$\gamma$ and $\alpha$, can be iterated so as to provide an
alternative to the sequential approach (for $\alpha >0$). The
iterations introduce a history-dependence and lead to the
emergence of $k_{\times}$, rendering both reconstruction methods
qualitatively similar.

For both approaches the new degree distributions have been
calculated and the degree-degree correlations emerging in the
depreciated network have been characterized, by means of standard
combinatorial methods. The main finding as regards the
correlations is that the sequential approach causes {\em
disassortative mixing} in the depreciated network when $\alpha >
0$.

For the simultaneous reconstruction approach, the exact (since
correlation-free) percolation threshold $f_c$ is derived for
central bias ($\alpha >0$) as a function of (non-integer) moments
of the degree distribution, for $\overline \gamma > 3$. On the
other hand, for $\overline\gamma < 3$ the exact finite-size
scaling law for the vanishing of $f_c$ is obtained. These results
fully demonstrate the validity of our exponent mapping
(\ref{mappy}) for central bias.

A {\em generating functions} approach is introduced for
degree-dependent edge percolation, extending previous work on
random percolation. This approach allows to obtain the size
distribution of finite clusters close to the percolation
transition as well as other critical properties. If the network
can be treated as a tree, which is valid for all finite clusters,
the generating functions satisfy self-consistency equations. We
have derived the extensions of these equations for
degree-dependent percolation, allowing for \textit{correlated}
networks, and have shown that they reduce to the original
equations provided no correlations are present. We have also
derived the criterion for the percolation threshold for
degree-dependent percolation and have shown that it reduces to the
familiar Molloy-Reed criterion when correlations are absent.
Further, our self-consistency equations reduce, for random
percolation, to equations frequently encountered in the
literature. Our generating functions formalism is new in the sense
that it extends known results on random percolation to biased
percolation which may involve correlated networks. In the
following, however, we draw further conclusions for the
statistical properties (including critical exponents) of {\em
uncorrelated} networks only.

Using the equivalence between the $q \rightarrow 1$ limit of the
Potts model and edge percolation, we have shown that critical
exponents for our biased percolation problem can be obtained from
the Potts model free energy by extending this equivalence to
inhomogeneous (edge-dependent) couplings in the Potts model and
edge-dependent removal probabilities in percolation. The
generating functions approach has been combined with the extension
of the Fortuin-Kasteleyn construction for the Potts model and with
{\em finite-size scaling} in order to extract the critical
exponents of the percolation transition, for uncorrelated
networks. We have found that the critical exponents are functions
of $\overline\gamma$, assuming that the degree distribution after
depreciation is governed by degree exponent $\overline\gamma$,
asymptotically for large degree. For $\overline\gamma$ we obtain
critical exponents that reduce to literature values of random
percolation simply by substituting $\overline\gamma \rightarrow
\gamma$. However, in the more delicate regime $2 < \overline\gamma
< 3$ this correspondence is not satisfied. A critical assessment
of this discrepancy is not given here, but left to future
scrutiny. We conclude that, in all cases, the only way in which
the bias exponent $\alpha$ enters in the critical exponents of the
percolation transition, is through the new degree exponent
$\overline\gamma$.

Furthermore, we have used numerical simulations to study the
properties of the network after depreciation and near the
percolation transition. We verified that robust networks can turn
fragile under centrally biased failure and that fragile networks
can turn robust under (sufficiently) strong peripherally biased
failure, using the sequential approach. Although correlations are
introduced in this approach, the results agree well with the
predictions for uncorrelated networks. Also the cross-over
behavior of the new degree distributions was tested and found to
agree well with the analytical expectations. As regards
correlations introduced by the sequential approach, we have been
able to verify the occurrence of disassortative mixing predicted
theoretically for $\alpha >0$.

The critical properties at percolation were checked by simulations
using the (iterated) simultaneous approach and also compared with
results obtained by simulations using the sequential approach.
Specifically, we have found that for biased percolation the
sequential and the (iterated) simultaneous approach give rise to
different results. In particular, the size of the giant cluster
predicted by the generating functions theory agrees very well with
the simulations for the (iterated) simultaneous approach.
Nevertheless, the differences are often small and we may conclude
that both methods are qualitatively similar. Finally, we have also
provided evidence for the theoretically expected appearance of
cross-over effects and degree-degree correlations for the
(iterated) simultaneous approach. Overall, we conclude that good
agreement has been found between simulations and theory.

\section{Appendix}
There exists a more formal way to prove that the degree
distribution in the diluted net satisfies
$\overline{P}(\overline{k})\propto
\overline{k}\,^{-\overline\gamma}$ for large degrees
$\overline{k}$ when central bias is applied.

We start from the degree distribution in the diluted network
$\overline{P}(\overline{k})$ which was calculated in
Eq.~\eqref{binomial}. For large values of $k$, i.e., $k \gg
k_{\times}$, and $\alpha>0$, the probability of retaining a node
of degree $k$ falls off as $\rho_k\propto k^{-\alpha}$; this is
valid using the sequential approach (see Eq.~\eqref{scalubl}), as
well as the simultaneous one (see Eq.~\eqref{marginalpaparel}).

If both $k\rho_k$ and $k (1- \rho_k)$ are large, the binomial
distribution can be approximated by a normal distribution with
mean $k\rho_k$ and variance $k\rho_k(1-\rho_k)$. The latter
condition is always true if we apply CB, since then $1-\rho_k
\approx 1$ for large $k$ as edges between the most connected nodes
are almost certainly removed.  Since $k \rho_k \propto k^{1 -
\alpha}$, the first requirement holds only if $\alpha <1$.

Inserting the normal distribution with variance $k\rho_k(1-\rho_k)
\approx C_0k^{1-\alpha}$, with $C_0$ a constant,  in
Eq.~\eqref{binomial} and approximating the sum by an integral
yields
\begin{align}
\overline{P}(\overline{k}) \propto \int_{\overline{k}}^{\infty}
dk\:k^{-\gamma+\frac{\alpha -
1}{2}}\exp\left(-\frac{\left[\,\overline{k}-C_0k^{1-\alpha}\right]^2}{2C_0k^{1-\alpha}}\right).
\end{align}
Now we introduce the auxiliary variable $u\equiv
k/\overline{k}^{\,1/(1-\alpha)}$ and rewrite the integral as
follows:
\begin{align}
\overline{P}(\overline{k})\propto
\overline{k}^{\frac{1-\gamma}{(1-\alpha)}-\frac{1}{2}}
&\int_{\overline{k}^{\frac{-\alpha}{1-\alpha}}}^{\infty}du\:
u^{-\gamma + \frac{\alpha-1}{2}}\nonumber
\\
&\times\exp\left(-\overline{k}\frac{\left[1-C_0u^{1-\alpha}\right]^2}{2C_0u^{1-\alpha}}\right).
\end{align}
For $\overline{k}\rightarrow \infty$, the integrand has only
non-vanishing values in a neighborhood $\Delta u
\approx1/\sqrt{\overline{k}}$ around $u_c = C_0^{-1/(1-\alpha)}$.
If $0<\alpha <1$, the lower bound of integration vanishes for
large $\overline{k}$. Thus $u_c$ certainly lies in the domain of
integration and the integral can simply be approximated by
$C_1/\sqrt{\overline{k}}$ with $C_1$ a constant. After some
trivial power counting, we arrive at
\begin{align}
\overline{P}(\overline{k}) \propto
\overline{k}\,^{-\frac{\gamma-\alpha}{1-\alpha}}. \label{gammaac}
\end{align}
Thus we obtain the anticipated behavior for CB. The exponent
$\overline{\gamma}$ controls the decay of the degree distribution
in the diluted network at large $\overline{k}$.

The case $\alpha = 1$ can be studied using a Poisson distribution
approximation for the binomial factors, which results in
$\overline{P}(\overline{k})$ being a Poisson-type degree
distribution $\overline{P}(\overline{k}) \sim
\overline{k}^{\,1-\gamma}/\overline{k}!$ for large $k$. Thus
$\alpha < 1$ is a natural restriction, because the scale-free
behavior is destroyed if stronger CB is applied.

We still have to examine the situation for $\alpha = 0$. Then the
lower bound of the integral becomes $1$, while $u_c = 1/C_0$. The
constant $C_0$ is nothing but the fraction of edges preserved
after the depreciation process. Thus $C_0 < 1$ and $u_c$ also lies
in the integration domain. Thus the previous arguments apply as
well to the random node removal process. We conclude that indeed
$\overline{P}(\overline{k})\sim
\overline{k}^{\,-\overline{\gamma}}$ for $0\leq\alpha<1$ which
proves the intuitive conjecture given in the Introduction.

\end{document}